# High-j Intruder Orbitals and the $\Delta I = 4$ Bifurcation in Superdeformed Bands

I. M. Pavlichenkov

*Russian Research Center "Kurchatov Institute", Moscow, 123182, Russia*

All the known cases of the $\Delta I = 4$ bifurcation in the SD bands are analysed in the framework of the microscopic theory suggested by the author [Phys. Rev. C **55** (1997) 1275]. It is shown that the intruder orbitals are of paramount importance to the energy staggering.

This talk consists of three parts: a review of the experimental situation; a short description of the theory; and its comparison with available experimental data.

1. One of the recent exciting discoveries in the superdeformed (SD) rotational bands was the regular staggering pattern in the transition energies of the yrast SD band in $^{149}$Gd [1]. The states of the $\Delta I = 2$ rotational band are split into two $\Delta I = 4$ subbands separated by about 100 eV. Since then such bifurcation (or the $\Delta I = 2$ staggering) has been found in the SD bands of $A=130$ [2] and $A=190$ [3] mass regions. There is also some tentative evidence of the staggering in the SD bands of $^{154}$Er [4] and $^{192}$Tl [5]. In the beginning, the phenomenon showed some symptoms of "Pathological Science." The effect was of a magnitude close to the limit of detectability and had a very low statistical significance. Fantastic theories contrary to experience have been suggested. There were claims of great accuracy.

Recently new measurements have been made with higher statistics using the Gammasphere multidetector array [6, 7]. The last work confirmed the original $^{149}$Gd long regular staggering pattern. Besides the $\Delta I = 4$ bifurcation has been observed in two new SD bands of $^{148}$Eu and $^{148}$Gd, which are identical to the yrast SD band of $^{149}$Gd. Both of these bands exhibit the same long and regular staggering pattern as the reference band. The length, regularity and striking similarity of the staggering pattern in these three bands refute the band crossing scenario [8].

It seems natural to explain the phenomenon by the static $Y_{44}$ deformation of a nuclear shape as suggested by Hamamoto, Mottelson [9] and Macchiavelli *et al.* [10]. However, an attempt to go farther on this way and understand the staggering from first principles had failed [11]. In another approach, suggested first in Ref. [1], the $\Delta I = 2$ staggering was considered to be a result of the $C_4$ symmetry perturbation. The phenomenological treatment accomplished in Ref. [12] confirmed the dynamic origin of the $C_4$ perturbation. The extension of this idea on a microscopic level was made in Refs. [13, 14]. These works are based on the mean field approach with the perturbation method, developed in the theory of nonadiabatic corrections to nuclear rotational spectra [15–17].

2. The starting point of the theory is the second quantized Routhian

$$H' = \sum_{i,k}\varepsilon_{ik}a_i^+ a_k - \frac{1}{2}\sum_{\lambda=2,4}\chi_\lambda \mathbf{Q}_\lambda \cdot \mathbf{Q}_\lambda - \omega_1 J_1 - \omega_2 J_2, \qquad (1)$$

with the quadrupole and hexadecapole residual interactions of strengths $\chi_2$ and $\chi_4$ respectively. The dot stands for the scalar product $\mathbf{Q}_\lambda \cdot \mathbf{Q}_\lambda = \sum_\mu Q^*_{\lambda\mu} Q_{\lambda\mu}$ of the multipole moment operators $Q_{\lambda\mu} = \sum_{i,k} q_{ik}(\lambda\mu) a_i^+ a_k$. The first term in Eq. (1) is the single-particle energy. The last two are cranking terms involving rotation around an *arbitrary* axis perpendicular to the symmetry axis 3. $J_s = \sum_{i,k} j_{ik}(s) a_i^+ a_k$ and $\omega_s$, $s=1,2$, are the projections of the angular momentum and velocity of a nucleus. For simplicity we ignore the pairing interaction.

Since all the numerical calculations done so far show the absence of any noticeable triaxial superdeformation, it is reasonable to confine ourselves to the axial case. Accordingly we perform the canonical transformation to the new single particle functions $|\nu\rangle$, which satisfy the self-consistent Schrödinger equation

$$(\hat{\varepsilon} - \sum_{\lambda=2,4} \chi_\lambda \alpha_{\lambda 0} \hat{q}_{\lambda 0} - \omega_+ j_- - \omega_- j_+) | \nu \rangle = \epsilon_\nu | \nu \rangle, \qquad (2)$$

where $j_\pm = j_1 \pm i j_2$, $\omega_\pm = \omega_1 \pm i \omega_2$, and $\alpha_{\lambda 0} = \sum_\nu q_{\nu\nu}(\lambda 0) n_\nu$ is the axial collective coordinate depending on the nucleon occupation number $n_\nu$. In the Hartree approximation, the energy in the rotating frame is given by

$$\mathcal{E}' = \sum_\nu \epsilon_\nu n_\nu + \frac{1}{2} \sum_\lambda \chi_\lambda \alpha_{\lambda 0}^2 - \frac{1}{2} \sum_{\lambda,\mu \neq 0} \chi_\lambda \alpha_{\lambda\mu} \alpha_{\lambda,-\mu}. \qquad (3)$$

The last term of this equation contains the nonaxial collective coordinates

$$\alpha_{\lambda\mu} = \text{Tr}(q_{\lambda\mu} \rho), \quad \mu = \pm 2, \pm 4, \qquad (4)$$

which are nonzero because the density matrix $\rho$ in the representation of Eq. (2) is not axial due to the cranking term. Since the nucleons in all orbitals except high-$j$ ones are only slightly disturbed by rotation, let us decompose the nonaxial collective coordinates into two parts

$$\alpha_{\lambda\mu} = Q_{\lambda\mu} + \tilde{\alpha}_{\lambda\mu}, \qquad (5)$$

where $Q_{\lambda\mu}$ is the multipole moment of the nucleons in the high-$j$ shells. Being intruder the states of a such unique-parity shell mix weakly with other single-particle states. This means that a filled intruder shell does not make a contribution to $Q_{\lambda\mu}$ because the operator $q_{\lambda\mu}$ has a vanishing track. Therefore, $Q_{\lambda\mu}$ is the multipole moment of the nucleons in the unfilled unique-parity shells near the Fermi surface.

The second part, $\tilde{\alpha}_{\lambda\mu}$, is the multipole moment of nucleons in the low-$j$ orbitals. It is possible to use the perturbation theory to obtain the $\omega$-dependence of this macroscopic quantity. Since the contribution of the intruder orbitals into the total nuclear multipole moment is small, $Q_{\lambda\mu} \sim \alpha_{\lambda\mu}/A$, we have:

$$\tilde{\alpha}_{\lambda\mu} \approx \text{Tr}(q_{\lambda\mu} \rho^{(2)}) + \text{Tr}(q_{\lambda\mu} \rho^{(4)}) + ..., \qquad (6)$$

where $\rho^{(n)}$ is the $n$th order correction to the nuclear density matrix due to the cranking term of Eq. (2). In performing corresponding calculations, we will use the representation of nonrotating axially-deformed nuclei. It is important to note that the perturbation series for the axial collective coordinates, $\alpha_{\lambda 0}$, involves the $\omega$-independent quantity $\alpha_{\lambda 0}^{(0)}$ (the multipole moment of nonrotating nucleus):

$$\alpha_{\lambda 0}^{(0)} = \text{Tr}(q_{\lambda 0}\rho^{(0)}). \tag{7}$$

The decomposition (5) and Eq. (6) allow us to obtain the expansion of the energy (3) in the powers of the rotational frequencies and then in the powers of the angular momentum components $I_{\pm} = I_1 \pm iI_2$ (see for details Refs. [14]). In order to find a condition for a given band to stagger, we consider the limit of *a pure collective rotation*. Then the Hamiltonian describing nonaiabatic effects in an isolated rotational band has, to the quartic terms in the angular momentum, the form:

$$H_{\text{eff}} = \mathcal{A}\mathbf{I}^2 + \mathcal{B}\mathbf{I}^4 + d(I_+^2 + I_-^2) + c(I_+^4 + I_-^4), \tag{8}$$

The expansion (8) contains axial and nonaxial terms. The axial ones include the moment of inertia $\Im$ ($\mathcal{A} = 1/2\Im$) and the next inertial parameter $B$ (see Ref. [15]). The nonaxial terms with the parameters

$$d = -\frac{1}{4\Im^2}\sum_{\lambda=2,4}\chi_\lambda Q_{\lambda 2}\left[\alpha_{\lambda 2}^{(2)} + \frac{I(I+1)}{4\Im^2}\alpha_{\lambda 2}^{(4)}\right], \quad c = -\frac{\chi_4}{16\Im^4}Q_{44}\alpha_{44}^{(4)}, \tag{9}$$

involve the quadrupole and hexadecapole coupling of the intruder states with the nonaxial distortion (due to the rotation) of the nuclear self-consistent field. The latter is described by the $\omega$-independent values

$$\alpha_{\lambda\mu}^{(n)} = \alpha_{\lambda,-\mu}^{(n)} = \omega_+^{-(n+\mu)/2}\omega_-^{-(n-\mu)/2}\text{Tr}(q_{\lambda\mu}\rho^{(n)}). \tag{10}$$

Let us note that the parameters $c$ and $d$ are the functions of $I/\Im$ because of the $\omega$-dependence of the values $Q_{\lambda\mu}$. The quadratic and quartic nonaxial terms become comparable for $I \sim I_c$, where

$$I_c = \sqrt{\frac{|d|}{4c}}. \tag{11}$$

The effective rotational Hamiltonian (8) is obtained by a $SU(2)$ mapping from the original fermion space to the rotor one. This method allows us to separate the rotational and the single-particle motion at the cost of nonadiabatic terms. For small spin $I$, when the values $Q_{\lambda\mu}$ are close to zero, Eq. (8) is reduced to the standard rotational Hamiltonian of axially-symmetric nuclei, the power series of the operator $\mathbf{I}^2$. This Hamiltonian does not lead to staggering. If a nucleus has a stable nonaxial deformation, the collective coordinates $\alpha_{22}$, $\alpha_{42}$, and $\alpha_{44}$ have $\omega$-independent parts by analogy with $\alpha_{\lambda 0}^{(0)}$ (see Eq.(7)). The nonaxial terms of such rotational Hamiltonian are $A$ times greater than the ones of axial nuclei. A similar Hamiltonian has been considered in Ref. [18], where it has been shown that it generates the irregular $\Delta I=2$ staggering. The suitable microscopic model has been studied in Ref. [19]. For the $Y_{44}$ deformation with respect to the symmetry axis, $\alpha_{44}$ is the only coordinate that involves an $\omega$-independent part. Accordingly, the Hamiltonian has the $C_{4v}$ symmetric form of Ref. [9]. In obtaining the Hamiltonian (8) we neglected the centrifugal effects due to collective vibrations since corresponding terms are proportional, not to $Q_{\lambda\mu}$, but to the coherent sums involving only a small fraction of intruder orbitals [16].

The Hamiltonian (8) is not fourfold invariant. The term with the operator $I_+^2 + I_-^2$ violates the $C_{4v}$ symmetry. However, it is invariant under the $D_2$ point symmetry group and under the transformation $d \to -d$ with the simultaneous rotation by 90° around the symmetry-axis. The later invariance allows us to consider only the negative values of $d$, in which case the yrast band corresponds to the rotation around the 1-axis if $I < I_c$. The energy $E_0(I)$ of the yrast states is determined in the quasiclassical approximation (see Ref. [20]) by the Bohr quantization condition:

$$S_1(E) = \int_0^{2\pi} I_1(\phi, E) d\phi = 2\pi I, \qquad (12)$$

where the $\phi$-conjugate variable $I_1$ is the angular momentum about the axis of quantization, i.e., the 1-axis. There are two sets of the equivalent classical trajectories surrounding the 1-axis on the rotation energy surface. This means that the energy levels of the band occur in degenerate pairs. The angular momentum tunneling splits the degeneracy leading to the $C_{2v}$ doublets. Only a single full symmetric state of the doublet is appropriate for the yrast SD band. Its energy is given by

$$E(I) = E_0(I) - 2 \mid T \mid \cos\{Re S_2(E_0)\}, \qquad (13)$$

where
$$T = \Big(\frac{\partial E}{\partial S_1}\Big)_{E_0} e^{i S_2(E_0)}, \ S_2(E) = \int_\gamma I_2(\phi, E) d\phi, \qquad (14)$$

is the tunneling amplitude for the angular momentum precessing on one trajectory to leak over on another. The tunneling path $\gamma$ connects the points on the equivalent trajectories where $I_2 = 0$ and forms a part of the great circle, which crosses the saddle point at the 3-axis. Since the angle $\phi$ is around the 2-axis the representation of the Hamiltonian (8) should be used with the 2-axis as the axis of quantization. Transforming the Hamiltonian to new axes we can estimate $S_2(E)$ for the classical minimum $E_{\min} = (\mathcal{A} + 2d)I^2 + (\mathcal{B} + 2c)I^4$. The action $S_2$ calculated in this way acquires a real part if $c > 0$ and $I_0 < I < I_c$, where

$$I_0 = (2 - \sqrt{2})^{1/2} I_c = \left[(2 - \sqrt{2})\frac{\mid d \mid}{4c}\right]^{1/2}. \qquad (15)$$

As spin $I$ increases in this region, the second term of Eq. (13) changes a sign every time the $Re S_2$ value increases by $\pi$. Thus, we have an irregular staggering. For the negative value of the parameter $c$, the staggering is absent in the lower levels of the I-multiplets (the yrast band), but it exists in upper levels because the transformation $c \to -c$ results in the inversion of multiplet levels. It should be noted that the staggering is independent of the sign of $c$ for the $C_{4v}$ invariant rotational Hamiltonian.

3. The above analysis shows that the staggering phenomenon is connected with the wave function oscillations in the classically forbidden regions of the rotational motion phase space. This effect is unusual for the second order Schrödinger equation and is the characteristic of a fourth order wave equation. That is why the parameter $c$ plays an important role. It involves the microscopic and macroscopic factors, $Q_{44}$ and $\alpha_{44}^{(n)}$. In order to calculate the former we use the eigenfunctions of Eq. (2), which, after performing the rotation of the intrinsic coordinate system by the angle $\eta$ around the 3-axis ($\cos\eta = \omega_1/\omega$, $\omega = \sqrt{\omega_1^2 + \omega_2^2}$), has the following form for an intruder $j$-shell:

$$\{\varepsilon_{nlj} + \kappa_2 j_3^2 - \kappa_4[7j_3^4 - (6\mathbf{j}^2 - 5)j_3^2] - \omega j_1\}|jkr\rangle = \epsilon_{jkr} \mid jkr\rangle, \qquad (16)$$

with
$$\kappa_2 = \frac{27\beta_2}{j(j+1)}\text{MeV}, \quad \kappa_4 = \frac{23\beta_4}{(j-1)j(j+1)(j+2)}\text{MeV},$$

where $\varepsilon_{nlj}$ is the energy of the $j$-orbitals in a spherical potential, $\beta_2$ and $\beta_4$ are the quadrupole and hexadecapole deformation parameters. The states of the $j$-shell are labeled by $k = 1, 2, ..., j + 1/2$ and the signature $r = \exp(-i\pi\alpha)$. The nonaxial hexadecapole moment of intruder orbitals is calculated with the help of the formula:

$$Q_{44} = \sum_{k,r} \langle jkr \mid q_{44} \mid jkr \rangle n_{jkr}, \qquad (17)$$

where $n_{jkr}$ are the occupation numbers of high-$j$ intruder orbitals near the Fermi surface: $i_{13/2}$ for protons and $j_{15/2}$ for neutrons in the 150 and 190 mass regions. Figure 1 shows the dependence of this value on the number $\mathcal{N}$ of occupied orbitals in the $j=15/2$ shell and the rotational frequency $\omega$.

Similar dependencies are observed for the $j=11/2$ and 13/2 shells. The function $F_j(\mathcal{N}) = Q_{44}/\langle nl|r^4|nl\rangle$ is generic for the intruder configurations of the SD bands providing $\mathcal{N}$ is small or large. This can be understood if we consider the limit of the angular momentum $\mathbf{j}$ alignment, for which the expectation value of $q_{44}$ is proportional to the Clebsch-Gordan coefficient $\langle jm, 40|jm\rangle$, with $m$ being the projection of $\mathbf{j}$ on the rotation axis 1. The coefficient is positive for $m = j$ and negative for $m \geq j - 1$, besides $\sum_{m=1/2}^{j}\langle jm, 40|jm\rangle = 0$.

The second factor, $\alpha_{44}^{(4)}$, has a fixed sign. In the axially-deformed oscillator potential with the frequencies $\omega_z$ along the symmetry axis and $\omega_\perp$ in the plane perpendicular to it, we have in the limit $\omega_\perp - \omega_z \gg \omega_\perp + \omega_z$:

$$\alpha_{44}^{(4)} = \sqrt{\frac{35}{32\pi}} \frac{9\hbar^2}{64M^2\omega_\perp^4\omega_z^2} \left(\frac{\omega_\perp + \omega_z}{\omega_\perp - \omega_z}\right)^4 (3\Sigma_{zz} - 12\Sigma_{\perp z} + 3\Sigma_{\perp\perp} - \Sigma_{\Lambda\Lambda}), \qquad (18)$$

where $M$ is nucleon mass and the quantities $\Sigma_{ik}$ are the sums of the bilinear combinations of the oscillator quantum numbers $n_\perp$, $n_z$, $\Lambda$ over occupied orbitals. The quasiclassical approximation gives us: $\Sigma_{\perp\perp} = 3/4\Sigma_{zz}$, $\Sigma_{\perp z} = 1/2\Sigma_{zz}$, $\Sigma_{\Lambda\Lambda} = 1/4\Sigma_{zz}$ for SD bands and $\Sigma_{\perp\perp} = 3\Sigma_{zz}$, $\Sigma_{\perp z} = \Sigma_{\Lambda\Lambda} = \Sigma_{zz}$ for ND bands. Thus the value (18) is negative and the sign of $c$ depends on that of $Q_{44}$.

The parameter $d$ is calculated analogously. Both parameters $c$ and $d$ allow us to estimate the minimal spin $I_0$, for which the $\Delta I = 2$ staggering may be present. We have found for the $A = 150$ mass region the following values: $I_0 \sim 400$ for the SD nuclei without pairing, $I_0 \sim 40$ for the SD nuclei with pairing, and $I_0 \sim 10$ for the ND nuclei with pairing.

We consider now a more realistic situation when the intrinsic configuration of a SD band has the dominating single-particle component $\pi 6^m \nu 7^n$ with $m$ and $n$ being the number of protons and neutrons occupying intruder states. By assuming according to Baranger and Kumar [21] that the mean square radius and the deformation are the same for protons and neutrons (this means that the ratio of protons to neutrons is the same inside a nucleus), we find for a nucleus with $Z$ protons and $N$ neutrons

$$c = \frac{\chi_4}{16\Im^4}\left[\left(\frac{2Z}{A}\right)^{2/3} Q_{44}(\pi) + \left(\frac{2N}{A}\right)^{2/3} Q_{44}(\nu)\right]\left\{\left(\frac{2Z}{A}\right)^{2/3} \alpha_{44}^{(4)}(Z) + \left(\frac{2N}{A}\right)^{2/3} \alpha_{44}^{(4)}(N)\right\}. \quad (19)$$

If $N$ is equal to $Z$, the last formula transforms to the parameter $c$ of Eq. (9). For the intruder configuration $\pi 6^m \nu 7^n$, the sign of $c$ is determined by the value

$$\mathcal{F}(m,n) = 15 F_{13/2}(m) + 19 F_{15/2}(n), \quad (20)$$

which depends also on the deformation and angular frequency. The inequality $\mathcal{F} > 0$ is the necessary condition for the $\Delta I = 2$ staggering to occur in the band.

Let us apply the criterion obtained above to the explanation of experimental data. The simplest case is the yrast SD band of $^{151}$Tb, $^{151}$Tb(1), having the configuration $\pi 6^3 \nu 7^2$. The neutron and proton moments $Q_{44}$ are negative and the staggering is definitely absent [1]. The values $Q_{44}(\pi)$ and $Q_{44}(\nu)$ have different signs in the case of the band $^{152}$Dy(1) with the configuration $\pi 6^4 \nu 7^2$. This configuration leads to the large negative $\mathcal{F}$-value (see Table I). That is why the staggering is not observed in this band [1].

The dominant configuration $\pi 6^2 \nu 7^1$ was assigned to a number of SD bands in $^{148,149}$Gd and $^{148}$Eu. As it follows from Table I, the neutron contribution to the value $\mathcal{F}(2,1)$ compensates for that of protons. In this case we have to take into account the contributions from other unfilled shells. Besides, the approximation of an isolated $j$-shell (16) is inapplicable. Nevertheless we can explain the experimental results of Ref. [7] qualitatively by considering $^{149}$Gd(1) as the reference band with the positive value of $c$. Then the identical band $^{148}$Er(1) with the configuration $^{149}$Gd(1)$\otimes(\pi[301]1/2, \alpha = -1/2)^{-1}$ exhibits the staggering because the moment $Q_{44}(\pi)$ changes little after removing a proton from the $s_{1/2}$ shell. The same holds for the band $^{148}$Gd(4) with the configuration $^{149}$Gd(1)$\otimes(\nu[411]1/2, \alpha = -1/2)^{-1}$ (we use the labeling of the $^{148}$Gd bands according to Ref. [22] and their interpretation according to Ref. [23]). On the other hand, no statistically significant staggering has been observed in the bands 1 and 5 of $^{148}$Gd, which differ from the $^{149}$Gd(1) one by a neutron hole in the states [651]1/2 and [642]5/2 with the signature $\alpha = +1/2$. Being favored these states make a noticeable contribution to the moment $Q_{44}(\nu)$. Thus, the removal of these neutrons may change the sign of $c$. As to the band $^{148}$Gd(6) having the configuration $^{149}$Gd(1)$\otimes(\nu[651]1/2)^{-2} \otimes (\nu[770]1/2, \alpha = 1/2)$ at low rotational frequency and $^{149}$Gd(1)$\otimes(\nu[770]1/2, \alpha = -1/2)^{-1}$ at high one, the absence of the staggering is explained in the approximation of the dominant configurations $\pi 6^2 \nu 7^2$ and $\pi 6^2 \nu 7^0$.

TABLE I. Values $\mathcal{F}(2,1)$ for the dominant configurations $\pi 6^2 \nu 7^1$ ($^{148,149}$Gd) and $\pi 6^4 \nu 7^2$ ($^{152}$Dy(1)). The deformation parameters are taken from Ref. [24].

| $\omega, MeV$ | $^{148}$Gd | $^{149}$Gd | $^{152}$Dy |
|---|---|---|---|
| 0.4 | −0.1087 | −0.1158 | −0.4190 |
| 0.6 | 0.0038 | −0.0179 | −0.2571 |
| 0.8 | 0.1454 | 0.1132 | −0.3290 |

There are some difficulties with the application of obtained criterion to $^{194}$Hg and the cerium isotopes. Bands 1, 2 and 3 of $^{194}$Hg have the same intruder configuration, $\pi 6^4 \nu 7^4$. Accordingly, $\mathcal{F}(4,4)$ has a large positive value for all frequencies and representative deformations. It is not clear why the staggering is observed in bands 2 and 3 and is not observed in the yrast band. The case of the cerium isotopes is more complicated. Having the intruder configuration $\pi 5^4 \nu 6^1$ and $\mathcal{F}(4,1) > 0$ the yrast band of $^{131}$Ce shows a staggering effect. The yrast band of $^{132}$Ce and bands 1,2 of $^{133}$Ce have the configuration $\pi 5^4 \nu 6^2$ and $\mathcal{F}(4,2) < 0$ for all frequencies. Nevertheless the staggering has been observed in Ref. [2]. This observation is in a contradiction with the necessary condition obtained above.

In summary, the cause of the $\Delta I = 2$ staggering has been found. The staggering is induced by the hexadecapole coupling of neutrons and protons occupying intruder orbitals with the nonaxial distortion of the nuclear mean field by rotation. By confining oneself to the pure collective rotation of nucleus we have found the necessary condition for the staggering in a band. All the evidences of staggered and nonstaggered bands in the $A$=150 mass region do not contradict with this condition. For the configuration $\pi 6^2 \nu 7^1$, the staggering is sensitive to the occupation of nonintruder orbitals.

The author is grateful to G. de France for giving the experimental results on the SD bands in $^{148}$Gd prior to publication. This work is supported in part by Russian Fund for Fundamental Research through Grant No. 96-02-16115.

**FIGURE CAPTION**

Fig. 1. Expectation value of the multipole moment $Q_{44}$ in the different single-particle configurations of the cranked single $j=15/2$ shell for different rotational frequencies $\omega=$ 0.4 MeV (circles), 0.6 MeV (squares ), 0.8 MeV (triangles), and the representative deformations $\beta_2$=0.622, $\beta_4$=0.041.

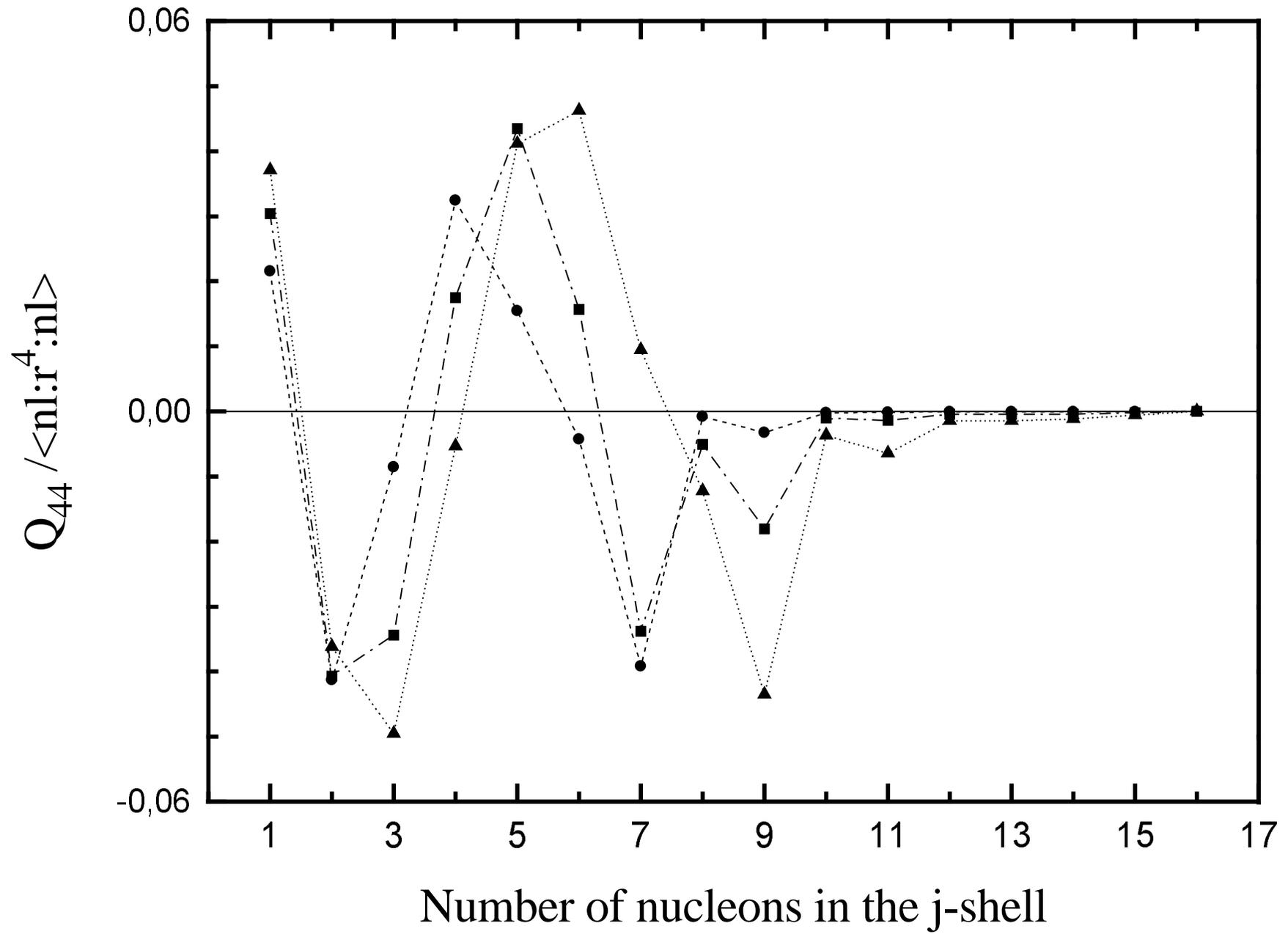